# Temperature behavior of optical absorption spectra in HfO$_2$ thin films

A.O. Shilov[1], S.S. Savchenko[1], A.S. Vokhmintsev[1], V.A. Gritsenko[2] and I.A. Weinstein[1,3]*

1 NANOTECH Centre, Ural Federal University, Yekaterinburg, Russia
2 Rzhanov Institute of Semiconductor Physics SB RAS, Novosibirsk, Russia
3 Vatolin Institute of Metallurgy UB RAS, Yekaterinburg, Russia

* Correspondence: i.a.weinstein@urfu.ru

**Abstract**

Hafnium dioxide, also known as hafnia, is an extremely sought-after material in opto- and nanoelectronics for creating optical coatings and various functional media to have stable performance characteristics under varying thermal operating conditions. In this paper, we have investigated the behavior of the optical properties of hafnia thin films exhibiting an amorphous structure in a wide temperature range of 7 - 296 K. For the first time we have examined the temperature effects in the energy gap of HfO$_2$ films and estimated the effective phonon energy of 30 meV responsible for observed thermally assisted shift of electronic levels. It has been shown that the electron-phonon interaction in the oxygen subsystem predominantly causes the observed changes. The obtained refractive index values for the tested films are established to be compatible with independent predicted data and to decrease as the temperature drops. The energy structure and electron-phonon interaction features that have been found are critical for forecasting how hafnia-thin-film-based optoelectronic devices will behave across a wide temperature range.

## 1. Introduction

HfO$_2$-based thin films have drawn interest from researchers around the world due to their high permittivity, optical transparency in the visible and near UV ranges, remarkable compatibility with silicon, excellent thermal stability, and other unique physicochemical properties [1–3]. Thin functional layers of hafnia that support devices such as MOSFET transistors, capacitors and optical coatings have significant potential for widespread application in micro-, opto-, and nanoelectronics [4–6]. Furthermore, considerable attention is currently focused on the possibilities of applying hafnia in new-generation devices in nanoelectronics, including memristive and ferroelectric memory cells [7–10], where the optical and electronic properties of the material play an important role.

As is well-known, the energy gap width $E_g$ of solids is one of the most important parameters that determine their intrinsic electronic and optical capabilities. It depends on external conditions (temperature, pressure, etc.), defect-phase composition, as well as the features of the energy structure and fundamental characteristics of the material's electron-phonon interaction [11,12]. As for HfO$_2$ nanostructures with different morphology and defect-phase composition, the literature primarily provides information on optical absorption at room temperature [13-21]. The circumstance concerned makes it difficult to accurately predict the behavior of the optical properties of hafnia thin films under varying thermal conditions. It also hinders the development of new and optimized electronic devices based on them that would exhibit the desired functional characteristics over a wide temperature range.

Taking into account the above-said, the aim of the paper is to experimentally study the temperature influence on the energy gap in thin-film hafnium dioxide and calculate the parameters of electron-phonon interaction. The regularities identified will enable one to describe more precisely the behavior of $HfO_2$ nanostructures at low temperatures, which can be utilized in designing modern opto- and nano-electronic devices.

2. **Materials and methods**

The work investigated $HfO_2$ thin films fabricated by the Rzhanov Institute of Semiconductor Physics SB RAS (Novosibirsk, Russia), utilizing ion-beam sputtering-deposition on a KU 9x9x1 $mm^3$ quartz substrate in an oxygen atmosphere. According to the manufacturer, the deposited films are amorphous, with a 300 nm thick oxide layer.

Measurements of optical absorption (OA) spectra in the wavelength range of 190 – 900 nm were taken using a Shimadzu UV-2450 spectrophotometer. For conducting the research at low temperatures, the samples were placed in a Janis CCS 100/204N cryostat equipped with a DT-670B-CU temperature sensor and a Model 335 controller. A HiCube 80 Eco pump was used to create a vacuum inside the cryostat (~ $7 \cdot 10^{-5}$ mbar). The absorption spectra were collected over two temperature ranges: 10–100 K with a step of 10 K and 100–296 K with a step of 20 K. Prior measurements of the quartz substrate's OA spectrum were made, and the absorption of $HfO_2$ films was analyzed using difference optical spectra.

3. **Results**

Figure 1 outlines the OA spectra measured over the temperature range of 7 – 296 K. It is seen that a sharp increase in absorption takes place at λ < 230 nm wavelengths, which corresponds to the intrinsic edge, whereas, at λ > 230 nm, the effects of light interference distort the OA spectrum. Upon cooling to cryogenic temperatures, the OA spectra shift towards the short-wavelength region, as in the inset to Figure 1. It should be noted that, in the range of 5.45–5.75 eV, a spectral shoulder arises against the backdrop of the OA edge.

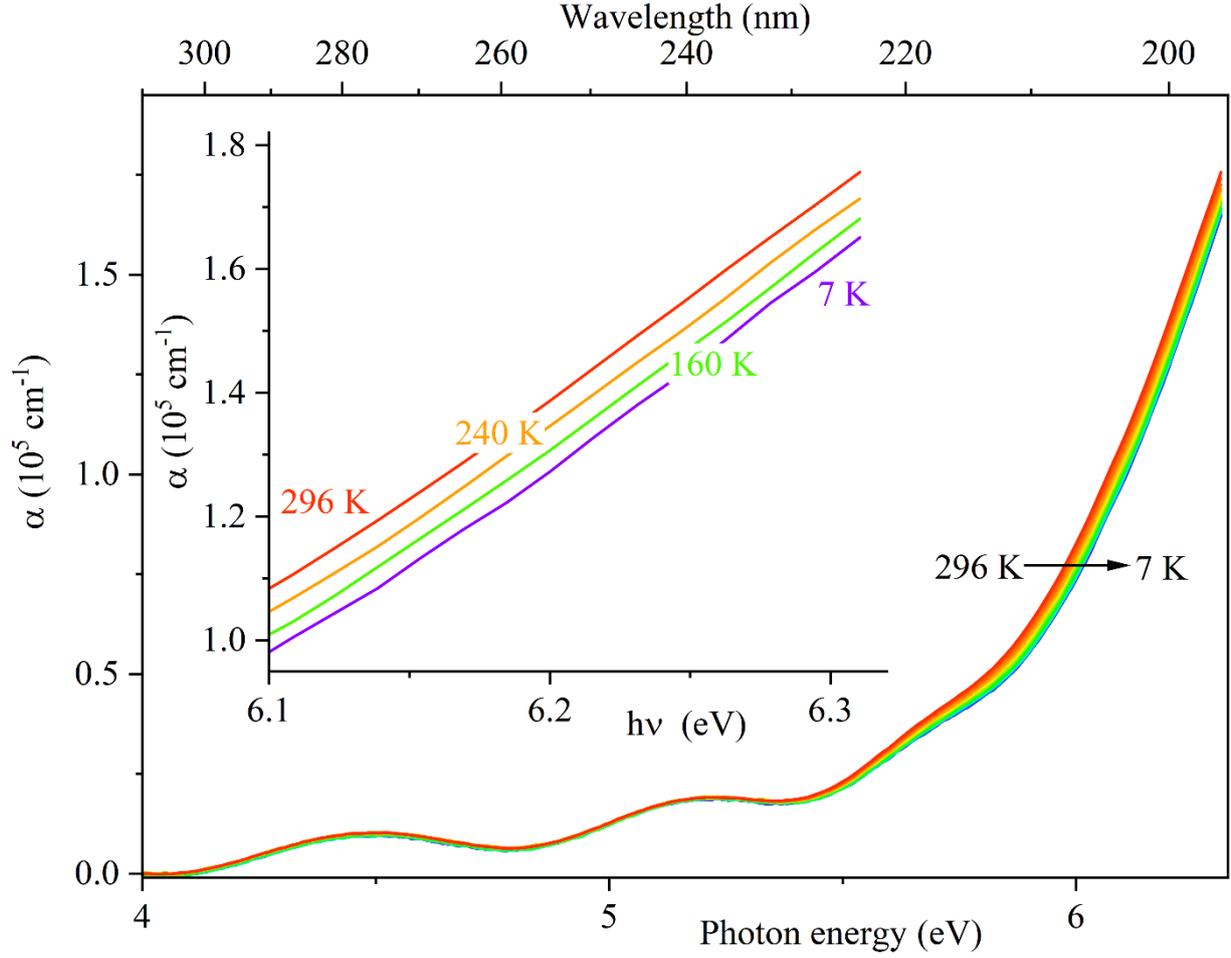

**Fig. 1** *OA spectra of HfO$_2$ thin films, measured within the temperature range of 7 – 296 K. The inset shows the short-wavelength segments of the OA spectra at different temperatures*

## 4. Discussion

It is worth emphasizing that we have previously analyzed the experimental data at room temperature [8], using a technique in Ref. [22]. The transformations performed allowed one to more accurately estimate the thickness of 288 ± 4 nm of the synthesized films, and calculate the spectral dependencies of the refractive index *n* and high-frequency permittivity $\varepsilon_\infty$ for HfO$_2$ within the range of 300 – 900 nm [8].

### 4.1 Evaluation of the energy gap width

The energy gap width $E_g$ in HfO$_2$, spanning the temperature range of 7 – 296 K, was estimated through the absorption coefficient α values acquired using the above-mentioned technique [22]. $E_g$ was determined using the well-known Tauc formalism [23]:

$$\alpha = \frac{B}{h\nu} \cdot (h\nu - E_g)^m. \qquad (1)$$

Here, *hν* is the incident photon energy, eV; *B* is the proportionality coefficient, eV$^{1-m}$ · cm$^{-1}$; *m* is the exponent depending on the type of optical transition.

Figure 2 shows the dependence $\alpha(h\nu)$ in the region of the optical absorption edge, measured at room temperature. Note that the spectral shoulder experimentally detected in the OA spectra at all the temperatures is not a manifestation of interference effects in thin films. Resorting to the known relations between the film thickness, the number of half-wavelengths and the previously established dependence $n(\lambda)$ [8], we found that a local maximum in the absorption spectrum, caused by destructive interference, should appear near 225 nm (5.5 eV). However, in our case, a local minimum of the spectrum is located in this region. Thus, it can be inferred that the interference effects in the strong absorption region impact negligibly. In addition, the analogous shoulder has been previously observed in the OA spectra of m-$HfO_2$ thin films [13,15] and nanoparticles of c-$HfO_2$ and m-$HfO_2$ [24]. For a quantitative analysis of the OA edge at hand, we used a superposition of the Tauc function for indirect transitions and a Gaussian-shaped band:

$$\alpha = \frac{B_1}{h\nu} \cdot (h\nu - E_g)^2 + \frac{B_2}{H\sqrt{\frac{\pi}{4\ln(2)}}} \cdot \exp\left(-\frac{4\ln(2)\cdot(h\nu - E_{max})^2}{H^2}\right). \quad (2)$$

Here, $B_1$, $eV^{-1} \cdot cm^{-1}$ and $B_2$, $eV \cdot cm^{-1}$ are the proportionality coefficients; $E_{max}$ is the position of the Gaussian peak maximum, eV, and $H$ is full width at half maximum (FWHM), eV.

In general, a Gaussian band can characterize optical transitions involving discrete levels in the energy structure of a material. Figure 2 exemplifies the OA spectrum approximation made for a temperature of 7 K using Eq. (2). It is evident that the exploited approach allows one to gain the experimental data such as $E_g$ (7 K) = 5.54 ± 0.01 eV, $E_{max}$ = 5.68 ± 0.02 eV, and $H$ = 0.30 ± 0.02 eV with high accuracy ($R^2 > 0.999$).

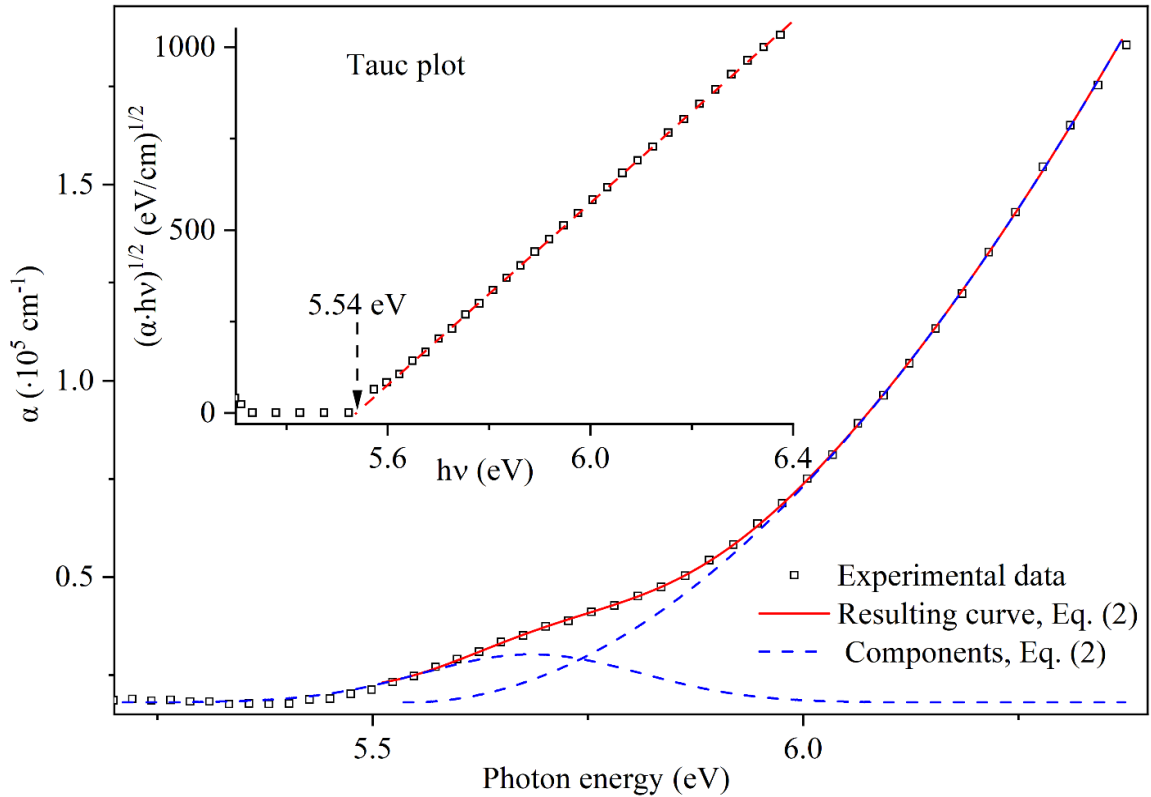

**Fig. 2** *$HfO_2$ OA spectrum approximation made for 7 K using a superposition of a power function for indirect interband transitions and a Gaussian. The inset shows the Tauc plot in appropriate coordinates and estimate of $E_g$.*

To correctly estimate the energy gap width using the Tauc approach, we removed the Gaussian shoulder from the OA spectrum to be then rebuilt in appropriate coordinates. The value of $E_g = 5.54$ eV was determined by intersecting the linear segment of the spectral dependence with the abscissa axis (a red dashed line), as seen in the inset in Figure 2. The values obtained are consistent with the known estimates of $E_g$ in $HfO_2$ with different morphologies, see Table 1. It is known that both direct and indirect allowed band-to-band transitions may occur in $HfO_2$ [13] [14, 25]. The current study points to the formation of the observed intrinsic absorption edge in the films involving indirect transitions. According to calculations in Ref. [25], $HfO_2$ structures with monoclinic crystal symmetry exhibit the above-specified transitions between the $\Gamma \rightarrow B$ points of the Brillouin zone.

**Table 1.** Energy gap width in $HfO_2$ thin films at room temperature

| Eg, eV | | Thickness, phase composition | Synthesis method | Reference |
|---|---|---|---|---|
| indirect | direct | | | |
| 5.49 ± 0.01 | — | 288 ± 4 nm amorphous | Ion-beam sputtering deposition | This work |
| 5.7 – 6.1 | — | 200-375 nm amorphous | Atomic layer deposition | [13] |
| 5.55 ± 0.03 | 5.68 – 5.72 | 200-375 nm monoclinic | | |
| 5.52 - 5.64 | 6.08 - 6.14 | 5 nm amorphous | | [14] |
| 5.50 ± 0.10 | — | 180-240 nm monoclinic | Radio-frequency sputtering deposition | [15] |
| — | 5.7 – 5.8 | 50-550 nm amorphous | | [16] |
| 5.54 ± 0.23 | — | 11-22 nm amorphous | | [17] |
| 5.53 ± 0.41 | — | 11-22 nm monoclinic | | |
| 5.49 ± 0.10 | — | 245-252 nm monoclinic | Electron-beam deposition | [18] |
| 5.54 ± 0.10 | — | 227 nm monoclinic | Ion-assisted deposition | |
| 5.6 | — | 80-95 nm amorphous | Spin coating | [19] |
| 5.13 | — | 76-81 nm monoclinic | | |
| 5.49 | — | 100 nm amorphous | Plasma-enhanced chemical vapor deposition | [20] |
| 5.5 | — | 15-50 nm amorphous | Atomic vapor deposition | [21] |

Obviously, the obtained energy values of $E_g$ are in good agreement with each other and with the results of other works [13-21]. The detected spectral shoulder of 5.68 eV in the region of the intrinsic absorption edge is discussed in works [13,15,18]. It is shown that this shoulder is typically indicative of amorphous films subjected to annealing, or a high-temperature transition to the monoclinic phase. Its position remains unaffected by the thickness of the oxide film.

**4.2 Temperature behavior of $E_g(T)$**

Using Eq.(2), we analyzed all the measured OA spectra of the studied films over the intrinsic edge region. Figure 3 displays the estimates of $E_g$ collected for different temperatures (black circles). It is known that an $E_g(T)$-temperature dependence for solids is caused by the electron-phonon interaction effects and can be quantitatively described via the Fan relation [26]:

$$E_g(T) = E_g(0) - A\left(e^{\frac{\hbar\omega}{kT}} - 1\right)^{-1}, \qquad (3)$$

where $E_g(0)$ is the energy gap width at a temperature of 0 K, eV; $A$ is the Fan parameter, which depends on the microscopic properties of a material, eV; $\hbar\omega$ is the energy of effective phonons responsible for the temperature shift of energy levels, eV. It should be noted that the parameters of the Gaussian shoulder remain almost unchanged over the entire temperature range. Figure 3 contains the calculated $E_g(T)$-dependence (a solid blue line) constructed exploiting Eq.(3). It mirrors the experimental data with a high degree of accuracy. The findings secured for the parameters in Eq.(3) are listed in Table 2. In addition, for the calculated function $E_g(T)$, we determined the temperature coefficient $\beta(T) = dE_g/dT$; see the solid red line in Figure 3. In the limit of high temperatures ($kT >> \hbar\omega$), one can write down [26]:

$$\beta_\infty = \frac{Ak}{\hbar\omega}. \qquad (4)$$

The value computed in this manner is represented in Table 2 and in Figure 3 by the dashed red line. It should be emphasized that, in the high-temperature limit, the value inferred for the coefficient $\beta_\infty$ is asymptotic.

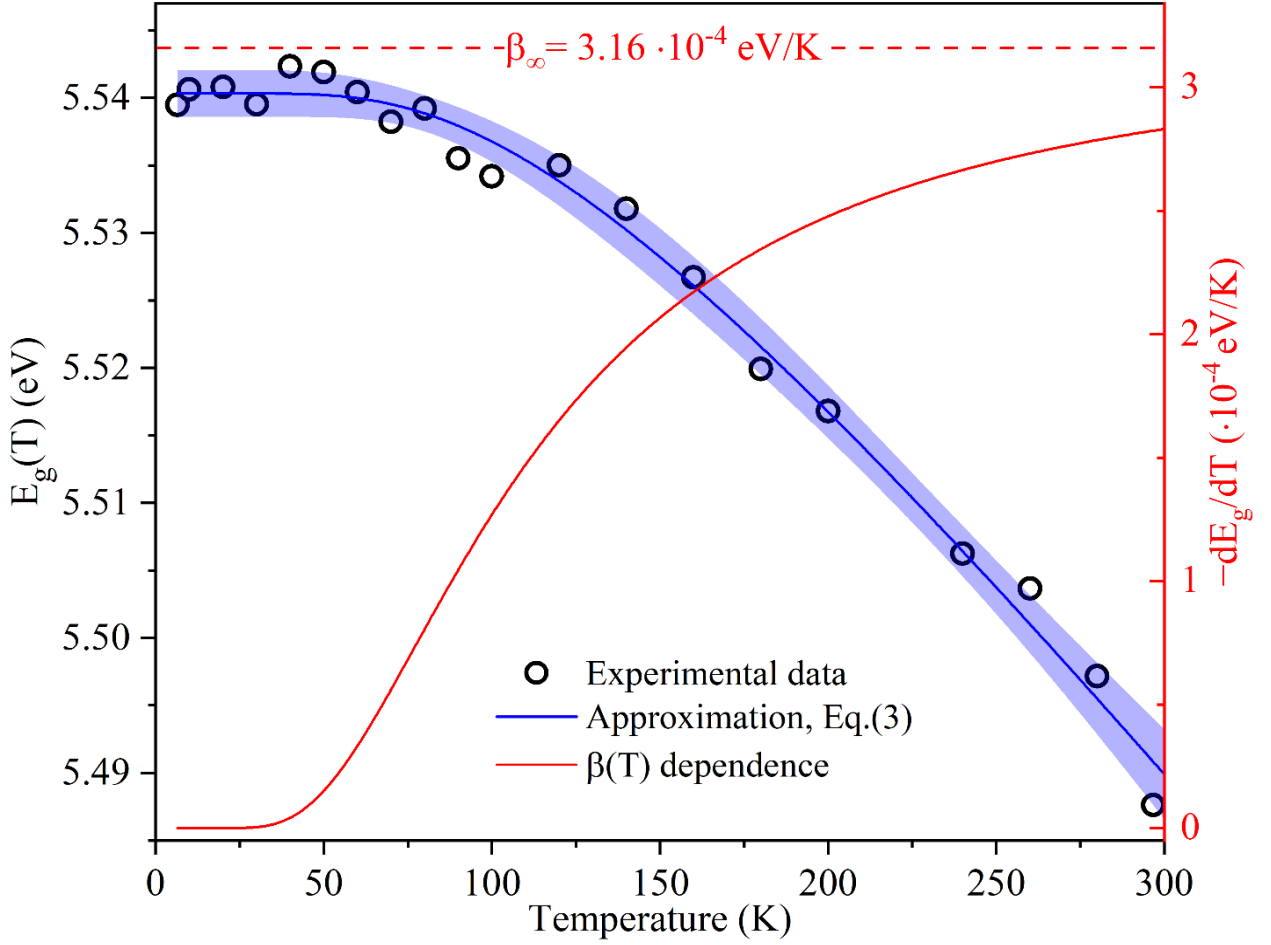

**Fig. 3** *Temperature dependence of the energy gap width. The blue curve is an approximation using the expression (3). The region shaded in blue covers the 99% confidence interval. The red solid line indicates the function β(T). The value of β∞ is shown by the dashed line.*

It is known that the ionic bonding with a degree of ionicity of 68% dominates in hafnium dioxide [27, 28]. Bearing in mind the statement, we can calculate the Fan parameter *A*, resorting to the expression applicable to ionic crystals [29]:

$$A = \frac{e^2}{\sqrt{2}\hbar}\sqrt{m_0 \hbar \omega}\frac{1}{4\pi\varepsilon_0}\cdot\left(\frac{1}{\varepsilon_\infty}-\frac{1}{\varepsilon}\right)\cdot\left[\left(\frac{m_e}{m_0}\right)^{\frac{1}{2}}+\left(\frac{m_h}{m_0}\right)^{\frac{1}{2}}\right], \qquad (5)$$

where $e$ is the electron charge; $m_0$ is the free electron mass; $\varepsilon_0$ is the electric constant; $\varepsilon_\infty$ and $\varepsilon$ are the high-frequency and static permittivities, respectively; and $m_e$ and $m_h$ are the effective masses of an electron and hole, respectively. To evaluate the parameter *A* in Eq.(5), we used independent data for monoclinic $HfO_2$, which are given with the corresponding references in Table 2. It is clear that Eq.(3) aligns this estimate closely with the Fan parameter value, assuming an approximation of the experimental data.

**Table 2.** Values of model parameters calculated or used in this paper

| Parameter | Value | Reference |
|---|---|---|
| Temperature behavior $E_g$ (T) | | |
| $E_g$ (0 K), eV | 5.54 ± 0.01 | This work, eq. (3) |
| $\hbar\omega$, meV | 30 ± 3 | This work, eq. (3) |
| | 32 ± 5 | [25] |
| Fan parameter A, eV | 0.11 ± 0.02 | This work, eq. (3) |
| | 0.103 | This work, eq. (5) |
| | 0.77 ± 0.23 | [25] |
| $\beta_\infty$, · 10$^{-4}$ eV/K | 3.16 | This work, eq. (4) |
| $\varepsilon$ | 25 | [30] |
| $\varepsilon_\infty$ | 4.345 | [8] |
| $m_e/m_0$ | 0.22 | [31] |
| $m_h/m_0$ | 0.17 | [31] |
| Cauchy model (6) | | |
| $n_0$ | 1.99 | This work, 296 K |
| | 1.93 | This work, 7 K |
| | 1.955 | [32] |
| | 1.875 | [33] |
| | 1.87–1.96 | [34] |
| $C_2$, nm$^2$ | 2.9 · 10$^3$ | This work, 296 K |
| | 1.6 · 10$^4$ | This work, 7 K |
| | 1.481 · 10$^4$ | [32] |
| | 6.28 · 10$^3$ | [33] |
| $C_4$, nm$^4$ | 9.5 · 10$^8$ | This work, 296 K |
| | 2.2 · 10$^{-6}$ | This work, 7 K |
| | 5.15 · 10$^8$ | [32] |
| | 5.8 · 10$^8$ | [33] |
| Penn model (7) | | |
| $\hbar\omega_p$, eV | 9.53 | This work, 296 K |
| | 9.06 | This work, 7 K |
| | 8.53 – 9.10 * | [34] |

*according to our estimates

We found that the value of $E_g$ increases by 0.05 eV as the temperature plummets from room temperature to 7 K. This circumstance agrees well with the fact of how the energy gap behaves in semiconductors and dielectrics, depending on temperature. In particular, similar temperature changes in the value of $E_g$ ranging from 0.05 to 0.18 eV are also observed in other broad-bandgap oxides and nitrides, including AlN and $Y_2O_3$ thin films [35,36]. It has been previously discovered [13] that, when cooled to a temperature of 10 K, photoluminescence excitation spectra of amorphous $HfO_2$ films and films with m-$HfO_2$ inclusions shift by 0.10 eV and 0.15 eV, respectively. Our approximation yields the energy of effective phonons as 30 meV, which is near that of the maximum at 300 cm$^{-1}$ (37 meV) in the IR spectra of m-HfO$_2$. This maximum is due to one longitudinal A$_u$ and two transverse B$_u$ vibrational modes of oxygen atoms [37]. The recent estimate of the energy of effective phonons at 32 meV, responsible for the shift and broadening of the exciton luminescence band, also meets the energy of the indicated vibrations [25]. Thus, it turns out that the characteristic values of $\hbar\omega$ almost coincide with the vibrational features of monoclinic hafnia. Consequently, the parameters of the short-range order and local atomic surrounding are close to each other. In addition, it can be concluded that the electron-phonon interaction involving 2p-orbital-oxygen-atom charge carriers as the valence band ceiling induces the temperature change in the energy gap width of the HfO$_2$ films [37].

The dispersion dependencies of the refractive index $n$ were calculated using a technique from Ref. [22] for different temperatures (see Figure 4) and then approximated within the Cauchy model [32,33]:

$$n(\lambda) = n_0 + \frac{C_2}{\lambda^2} + \frac{C_4}{\lambda^4}, \qquad (6)$$

where $n_0$ is the static refractive index as $\lambda \to \infty$; $C_2$ and $C_4$ are the fitting parameters of the proper dimensions. All the calculated values are given in Table 2. They are in good agreement with independent literature data [32,33].

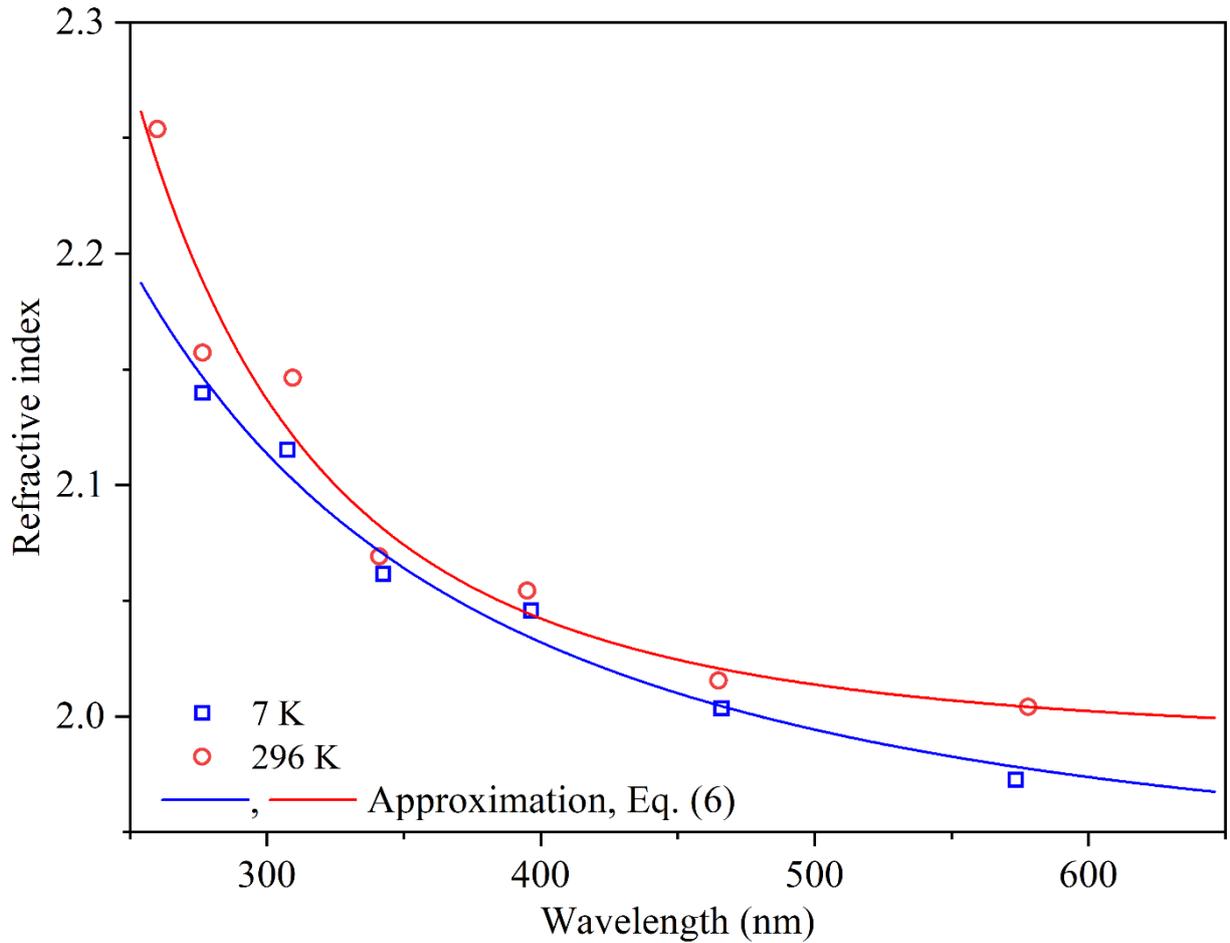

**Fig. 4** *Comparison of the dispersion dependencies of the refractive index for thin-film HfO$_2$, calculated based on the OA spectra at room temperature (red squares) and at 7 K (blue circles). Solid lines are approximations of the experimental data using the Cauchy model (6).*

It is evident that the refractive index rises as the temperature goes up, as is typical of wide-gap oxides [38]. It is worth pointing out that, the parameters $n$ and $E_g$, generally, characterize the material's electronic subsystem through a relationship within the Penn model [39,40]:

$$n_0^2 = 1 + \left(\frac{\hbar\omega_p}{E_g}\right)^2, \qquad (7)$$

where $\hbar\omega_p$ is the plasma electron energy that quantitatively determines the collective oscillation frequency of all valence electrons in the material, eV.

Table 2 summarizes the estimated model parameters. The values $n_0 = 1.93 - 1.99$ correlate well with independent data for monoclinic hafnia [34], evidencing the closeness of the local atomic surrounding parameters for amorphous and monoclinic $HfO_2$ nanostructures. The estimates of $\hbar\omega_p$, see Table 2, are quite consistent with the low-energy shoulder in the electron energy loss spectra (7-10 eV) due to the O 2p → Hf 3d transitions of valence electrons to the free states of hafnium d-orbitals [41–43]. According to eq. (7) $\hbar\omega_p$ increases with the temperature, see Fig.4 also. The same behavior is observed, for example, in thin films of indium phosphide [44].

**Conclusion**

In this work, the optical absorption spectra of hafnia thin films were studied in the temperature range of 7 – 296 K. It was found that the measured spectral dependences in the region of the intrinsic absorption edge can be approximated with a high accuracy by a superposition of a Gaussian band with a maximum of 5.68 ± 0.02 eV and a Tauc dependence for indirect allowed band-to-band transitions. The effect of temperature on the energy gap $E_g$ (T) of hafnium dioxide was analyzed for the first time in the range of 7–296 K. The parameters of dynamic disorder were estimated: $E_g(0) = 5.54 \pm 0.01$ eV, the effective energy of phonons $\hbar\omega = 30 \pm 3$ meV responsible for the shift of electronic levels, the temperature coefficient $\beta = 3.16 \cdot 10^{-4}$ eV/K. It was found that the decrease of the energy gap in $HfO_2$ films is due to the electron-phonon interaction involving electrons localized in the 2p orbitals of oxygen atoms that form the top of the valence band.

The temperature dependence of the refractive index was also studied. Using the Penn model the plasma energy was estimated $\hbar\omega_p = 9.53$ eV at 296 K and 9.06 eV at 7 K. The obtained results allow to more accurately predict the behavior of $HfO_2$ nanostructures at low temperatures, that is necessary for the development of modern opto- and nanoelectronic devices operating in wide temperature ranges.

**Acknowledgments**
The research funding from the Ministry of Science and Higher Education of the Russian Federation, project FEUZ-2023-0014, is gratefully acknowledged.

**Statements & Declarations**
**Conflict of interest:** The authors declare that they have no known competing financial interests or personal relationships that could have appeared to influence the work reported in this paper.